\documentclass{PoS}
\usepackage{amsmath}
\usepackage{amssymb}
\usepackage{url}
\usepackage{sidecap}
\usepackage[utf8]{inputenc}

\title{Up and down quark masses and corrections to Dashen's theorem from lattice QCD and quenched QED}

\ShortTitle{Up and down quark masses and corrections to Dashen's theorem}

\author{ \speaker{L. Varnhorst}$^{ag}$, S. Durr$^{ab}$, Z. Fodor$^{abc}$, C. Hoelbling$^{a}$, S. Krieg$^{ab}$, L. Lellouch$^{d}$, A. Portelli$^{def}$, A. Sastre$^{ad}$,                                                                                                                                                                                                                                                                                                                                                                                                                                                                                                                                                                                                                                                                                                                                                                                                                                                                                                                                                                                                                                                                                                                                                                                                                                                                                                                                                                                                                                                                                                                                                                                                                                                                                                                                                                                                                                                                                                                                                                                                                                                                                                                                                                                                                                                                                                                                                                                                                                                                                                                                                                                                                                                                                                                                                                                                                                                                                                                                                                                                                              K. K. Szabo$^{ab}$\\
	\llap{$^a$} Department of Physics, University of Wuppertal, Gaussstr. 20, D-42119 Wuppertal, Germany\\
	\llap{$^b$}IAS/JSC, Forschungszentrum Jülich, D-52425 Jülich, German \\
	\llap{$^c$}Institute for Theoretical Physics, Eötvös University, Pázmány P. sét. 1/A, H-1117 Budapest, Hungary\\
	\llap{$^d$}CNRS, Aix-Marseille U., U. de Toulon, Centre de Physique Théorique, UMR 7332, F-13288 Marseille, France\\
	\llap{$^e$}School of Physics \& Astronomy, University of Southampton, SO17 1BJ Southampton, United Kingdom\\
	\llap{$^f$}School of Physics \& Astronomy, The University of Edinburgh, EH9 3FD Edinburgh, United Kingdom\\
	\llap{$^g$}E-mail: \email{l.varnhorst@t-online.de}}

\abstract{We present a determination of the corrections to Dashen's theorem and of the individual up and down quark masses from a lattice calculation based on quenched QED and $N_f=2+1$ QCD simulations with 5 lattice spacings down to 0.054 fm. The simulations feature lattice sizes up to 6 fm and average up-down quark masses all the way down to their physical value. For the parameter which quantifies violations to Dashens's theorem we obtain $\epsilon=0.73(2)(5)(17)$, where the first error is statistical, the second is systematic, and the third is an estimate of the QED quenching error. For the light quark masses we obtain, $m_u=2.27(6)(5)(4) \, \mbox{MeV}$ and $m_d=4.67(6)(5)(4) \, \mbox{MeV}$ in the $\overline{\mbox{MS}}$ scheme at $2 \, \mbox{GeV}$ and the isospin breaking ratios $m_u/m_d=0.485(11)(8)(14)$, $R=38.2(1.1)(0.8)(1.4)$ and $Q=23.4(0.4)(0.3)(0.4)$. Our results exclude the $m_u=0$ solution to the strong CP problem by more than 24 standard deviations.}

\FullConference{34th annual International Symposium on Lattice Field Theory\\
		24-30 July 2016\\
		University of Southampton, UK}

\begin{document}

\section{Introduction}
The main contribution to hadron masses is due to the energy associated with the nonperturbative interactions of QCD while only a tiny part of the mass comes from the quark masses themselves. Due to the confining nature of the QCD interactions it is not possible to determine the quark mass parameters directly by experiment. A possible way to infer the correct values is to perform lattice QCD calculations in which correlation functions and masses of hadrons can be determined nonperturbatively from first principles. Matching these masses to the values observed in experiments allows for an ab-initio determination of the quark masses. The value of the average up and down quark mass $m_{ud}$ has been studied extensively both by the BMW collaboration \cite{Durr:2010aw,Durr:2010vn} and many other groups. A review of previous determination can be found in the FLAG report \cite{Aoki:2016frl}.

As in nature the up and down quark mass are not degenerate it is interesting to study their mass difference $\delta m = m_u - m_d$. The challenge in the computation of this quantity is not only that it requires non-degenerate quark masses in the lattice calculation but that the effect of the quark mass splitting to hadron mass is of order $\mathcal O(\delta m / \Lambda_{QCD})\approx 1\%$. This is not only very small but also comparable in magnitude to a different effect contributing to the hadron mass splitting namely the electromagnetic splitting of order $\mathcal O(\alpha)$. Therefore the electromagnetic interactions have to be included in the calculation for a reliable determination of $\delta m$. Related determinations can be found in \cite{Durr:2010aw,Durr:2010vn,Bazavov:2009fk,Duncan:1996xy,Blum:2007cy,Carrasco:2014cwa,Aoki:2012st,Aubin:2004ck,deDivitiis:2011eh,Blum:2010ym,deDivitiis:2013xla,Horsley:2015vla,Basak:2016jnn}. A review of the available results can be found in \cite{Aoki:2016frl}.

In this work the QED effects are treated in a quenched setup on top of the $N_f=2+1$ configurations used in \cite{Durr:2010aw}. These configurations feature pion masses all the way down to the physical point and allow a reliable continuum and finite-volume extrapolation. A determination of the isospin splitting of hadron masses in full QCD+QED can be found in \cite{Durr:2010aw}. However for a reliable determination of $\delta m$, lattice data down to physical pion masses is required \cite{Durr:2013goa} and hence the $N_f=2+1$ dataset was chosen for this analysis.

This lattice conference contribution summarizes an analysis which was originally published in \cite{Fodor:2016bgu} and is laid out as follows: 
After this introduction the lattice setup and the configurations used are described. We end with a presentations of the results achieved.

\section{Lattice setup}
The $N_f=2+1$ QCD configurations used in this study where generated with a tree-level $\mathcal O(a)$ improved Wilson fermion action with two steps of HEX smearing. As a gauge action we used a Symmanzik improved action. Details about the QCD configurations can be found in \cite{Borsanyi:2013lga}.

The QED effects are added on top of theese configurations by generating $U(1)$ gauge fields for each of the QCD configurations distributed according to a non-compact Maxwell action in Coulomb gauge. In the QED action the four-momentum zero mode was fixed to 0. This prescription is called $\mbox{QCD}_{\mbox{TL}}$ in \cite{Borsanyi:2014jba}. As by the introduction of QED effects the up and down quark masses renormalize differently this setup has to be considered as partially quenched.

On each resulting $SU(3)\times U(1)$ configurations two sets of propagators where calculated: In the first set the valence quark masses where tuned in a way that the resulting bare PCAC quark masses where equal to the bare PCAC quark masses on the same QCD configuration without QED effects. In the second set the bare $m_u$ and $m_s$ masses where set to the same values as in the first set while $m_d$ was varied to generate a spread in $\delta m$. On one particular QCD ensemble three set of propagators where generated. In two of them $\delta m$ was kept close to the physical value while the electromagnetic coupling was set to two or four times the physical value. In the third one the parameter where set so that $\delta m \approx 0$ and $\alpha \approx 0$.

\section{Analysis procedure}
To determine the splitting of the light quark masses one has to tune 5 parameters: $\alpha_s$, $\alpha$, $m_u$, $m_d$ and $m_s$. To define the physical point the following input quantities where employed: $M_{\pi^+}^2$, $M_{K_\chi}^2 = (M_{K^+}^2+M_{K^0}^2-M_{\pi^+}^2)/2$, the kaon mass splitting $\Delta M_K^2 = M_{K^+}^2 - M_{K^0}^2$ and the electromagnetic coupling in the Thompson limit. Additionally the lattice spacing has to be fixed by fitting the $M_\Omega^-$ or $M_{\Xi}$ mass. The kaon splitting was interpolated to the physical $M_{\pi^+}^2$, $M_{K_\chi}^2$ and $\alpha$ values using the leading order expansion
\begin{equation}
 \Delta M_K^2 = C_K(M_{\pi^+}^2, M_{K_\chi}^2, a, L) \alpha + \tilde D_K(M_{\pi^+}^2, M_{K_\chi}^2, a) \delta m
\end{equation}
where the first term on the right hand side is the electromagnetic contribution to the splitting and the second term is the splitting caused by non-degenerate quark masses. To avoid dealing with the complicated renormalization of $\delta m$, the leading order relation from partially quenched chiral perturbation theory with QED (PQ$\chi$PT+QED)
\begin{equation}
 \Delta M^2 = M_{\bar uu}^2 - M_{\bar dd}^2 = 2B_2\delta m + \mathcal O(m_{ud}\alpha, m_{ud}\delta m, \alpha^2, \alpha\delta m, \delta m^2)  \label{eqn_del_m}
\end{equation}
was used where $M_{\bar uu}$ and $M_{\bar dd}$ are the masses of the respective ``connected pseudoscalar mesons'' \cite{Borsanyi:2013lga}. The value of the low energy constant $B_2$ was determined in \cite{Durr:2013goa}. Using this relation
\begin{equation}
 \Delta M_K^2 = C_K(M_{\pi^+}^2, M_{K_\chi}^2, a, L) \alpha + D_K(M_{\pi^+}^2, M_{K_\chi}^2, a) \Delta M^2. \label{eqn_central_fit}
\end{equation}
can be derived.
Fitting this relation to the lattice data and using the physical value of $\Delta M_K^2$ allows to read off the value of $\Delta M$ at the physical point. Once this has be determined one can read off $C_K$ and $D_K$ at the physical point which allows to determine both the electromagnetic contribution to the Kaon splitting and the QCD contribution separately. Furthermore, once $\Delta M$ is determined, one can use the relation (\ref{eqn_del_m})
to extract the light quark mass splitting. Having at hand the value of the electromagnetic splitting of the Kaon mass one can determine the violation to Dashen's theorem: the electromagnetic mass splittings of the mesons in the pesudoscalar octet in the SU(3) flavor symmetric limit fulfill Dashen's theorem \cite{Dashen:1969eg}.
As nature features non-degenerate light and strange quark masses this theorem is violated. The strength of this violation can be parametrized by the parameter
\begin{equation}
 \epsilon = \frac{\Delta_{\mbox{QED}} M_K^2-\Delta_{\mbox{QED}} M_\pi^2}{\Delta M_\pi^2}. \label{eqn_eps}
\end{equation}
To evaluate this quantity one has in principle to determine $\Delta M_\pi^2$ on the lattice. However one can show using G-parity that to leading order in $\delta m$ the relation $\Delta M_\pi^2=\Delta_{\mbox{QED}} M_\pi^2$ is fulfilled and hence one can plug in the experimentally measured pion mass splitting to get both $\Delta M_\pi^2$ and $\Delta_{\mbox{QED}} M_\pi^2$.

\begin{SCfigure}[50]
 \centering
 \includegraphics[width=0.4\textwidth]{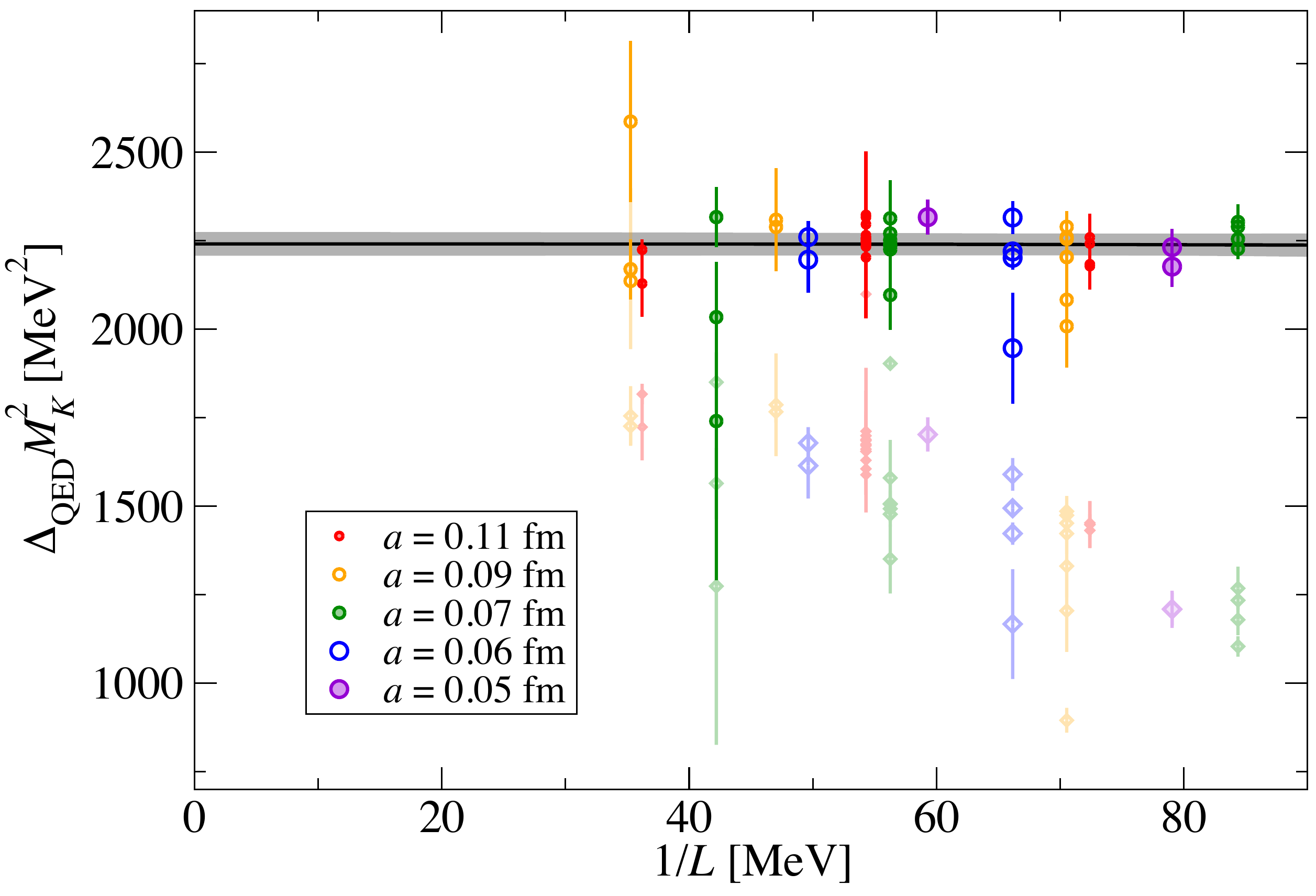}
 \caption{The Kaon splitting with and without the subtraction of the universal part of the finite volume corrections. The faded points are uncorrected and the solid points are corrected. The black line is a fit to the remaining finite volume dependence with a $L^{-3}$ ansatz. The gray band is the statistical error. All points have been projected to the physical point in all but the finite volume direction by the fit function. \label{fig_fvol}}
\end{SCfigure}

The nature of finite volume effects is very different for QCD and QED. In QCD there is a mass gap and therefore masses in finite volume receive a contribution that is exponentially suppressed with the extend of the system. QED however possesses no mass gap and hence there is no exponential suppression of the finite volume effects. Also the nature of the finite volume effects depend on the exact choice of the zero mode subtraction in the QED action \cite{Borsanyi:2014jba}. The corrections for various particles in $\mbox{QCD}_{\mbox{TL}}$ can be worked out \cite{Borsanyi:2014jba, Davoudi:2014qua} and the first two orders for charged scalar particles are not structure dependent and read
\begin{equation}
  \frac{M(L)}{M(\infty)} = - \frac{\kappa}{M(\infty)L} \left[ 1 + \frac{2}{M(\infty)L} \left( 1 - \frac{\pi}{2\kappa} \, \frac{T}{L} \right) \right]
\end{equation} 
where $\kappa=2.837\ldots$ is a known constant. One should note that there are two dimensionless quantities going into these equation: the mass of the particle times the spatial extend of the system and $T/L$. This is a feature of $\mbox{QCD}_{\mbox{TL}}$. For each charged particle this correction was applied prior to any fitting. The difference between the original data and the corrected data can be seen in figure \ref{fig_fvol}.

\begin{figure}
 \centering
 \includegraphics[width=0.8\textwidth]{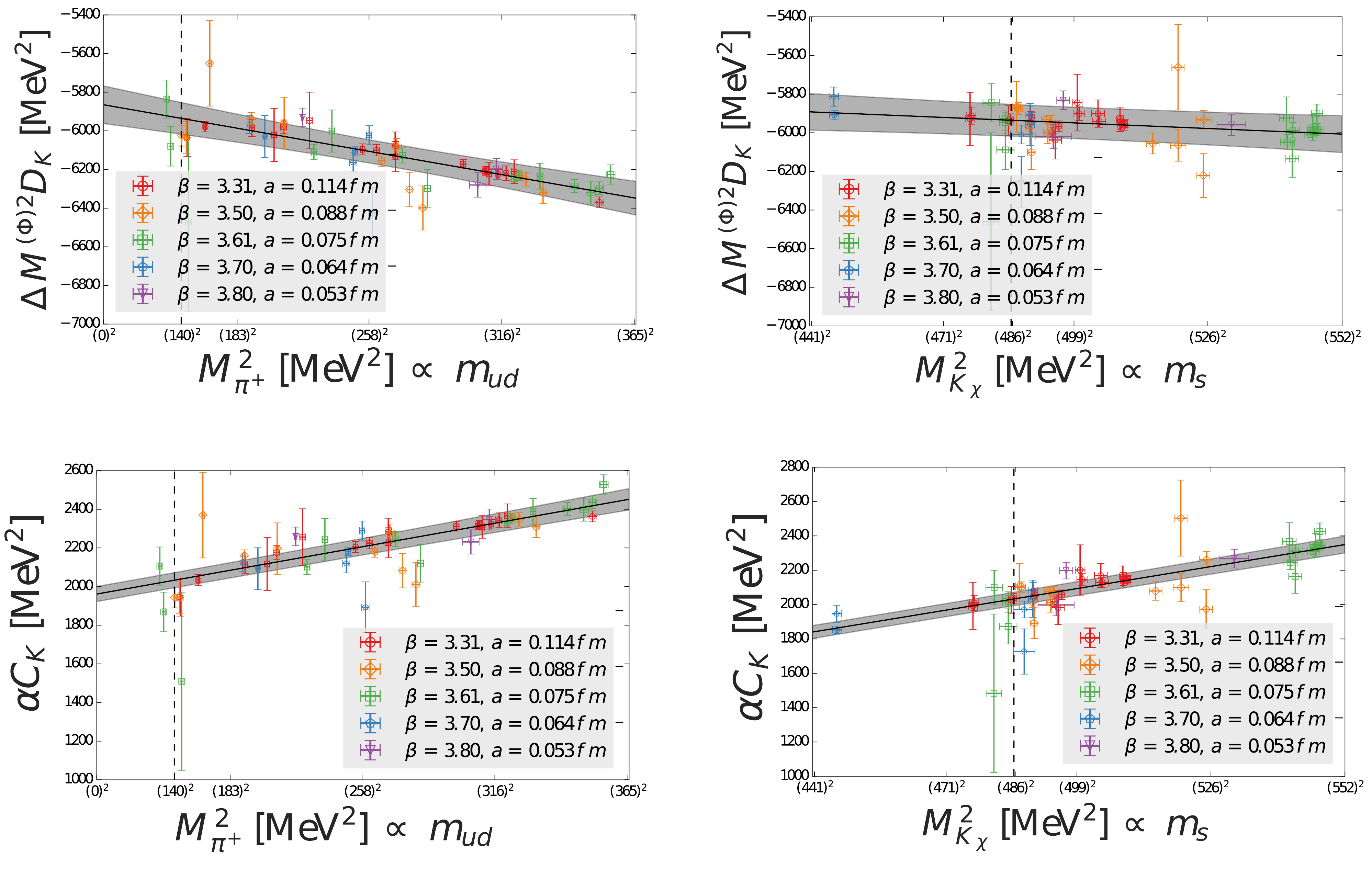}
 \caption{The $M_{\pi^+}^2$ and $M_{K_\chi}^2=0.5(M_{K^+}^2 + M_{K^0}^2 - M_{\pi^+}^2)$ behavior of $C_K$ and $D_K$ from one representation fit. All points have been projected to the physical point in all but the plotted direction by the fit function.\label{fig_fit}}
\end{figure}

To use eqn. (\ref{eqn_central_fit}) one has to find a suitable parametrization for $C_K$ and $D_K$ to interpolate the lattice data to the physical point. The following two fit ans\"atze were employed:
\begin{subequations}
\begin{align}
 C_K &= c_0 + c_1 (M_{\pi^+}^2 -M_{\pi^+}^{(\Phi)2})  + c_2 (M_{K_\chi}^2 -M_{K_\chi}^{(\Phi)2}) +c_3a + c_4 \frac{1}{L^3}, \label{eqn_CK}\\
 D_K &= d_0 + d_1(M_{\pi^+}^2 -M_{\pi^+}^{(\Phi)2}) + d_2 (M_{K_\chi}^2 -M_{K_\chi}^{(\Phi)2}) +d_3 f(a). \label{eqn_DK}
\end{align}
\end{subequations}
where $f(a)$ can be either $a^2$ or $\alpha_s a$ and $M_X^{(\Phi)}$ is the mass of $X$ at the physical point. A fully correlated fit of eqn. (\ref{eqn_central_fit}) using eqns. (\ref{eqn_CK}) and (\ref{eqn_DK}) to the Kaon splitting data was performed. The result of such a fit can be found in figure \ref{fig_fit}.

To estimate the systematic error the histogram method was employed \cite{Durr:2008zz}. Here the following variants of the fit functions where considered. Correlators where fitted with a conservative or aggressive time range, the scale was set either with the $\Omega^-$ or the $\Xi$ mass. In the scale setting all points with $M_\pi$ larger then either $400 \, \mbox{MeV}$ or $450 \, \mbox{MeV}$ where eliminated and in the fit to the $\Delta M_K^2$ data all points with $M_\pi$ larger then either $350 \, \mbox{MeV}$ or $400 \, \mbox{MeV}$ where eliminated. In $D_K$ either $a^2$ or $a\alpha_s$ contributions where considered. Finally in the expansions of $C_K$ and $D_K$ the Taylor expansions where replaced by Padé expansions independently. This resulted in a set of 128 different analyses. These analyses where weighted by their fit quality and a histogram was constructed. The spread off this histogram was taken as the systematic error. The statistical error was estimated using a bootstrap procedure. The QCD quenching uncertainty was conservatively estimated by using large $N_c$ counting and $SU(3)$ flavor symmetry considerations to be $\mathcal O(10 \%)$ on the $\mathcal O(\alpha)$ contribution to a given isospin splitting \cite{Borsanyi:2013lga}.

\section{Results}
\begin{figure}
\centering
\includegraphics[width=0.45\textwidth]{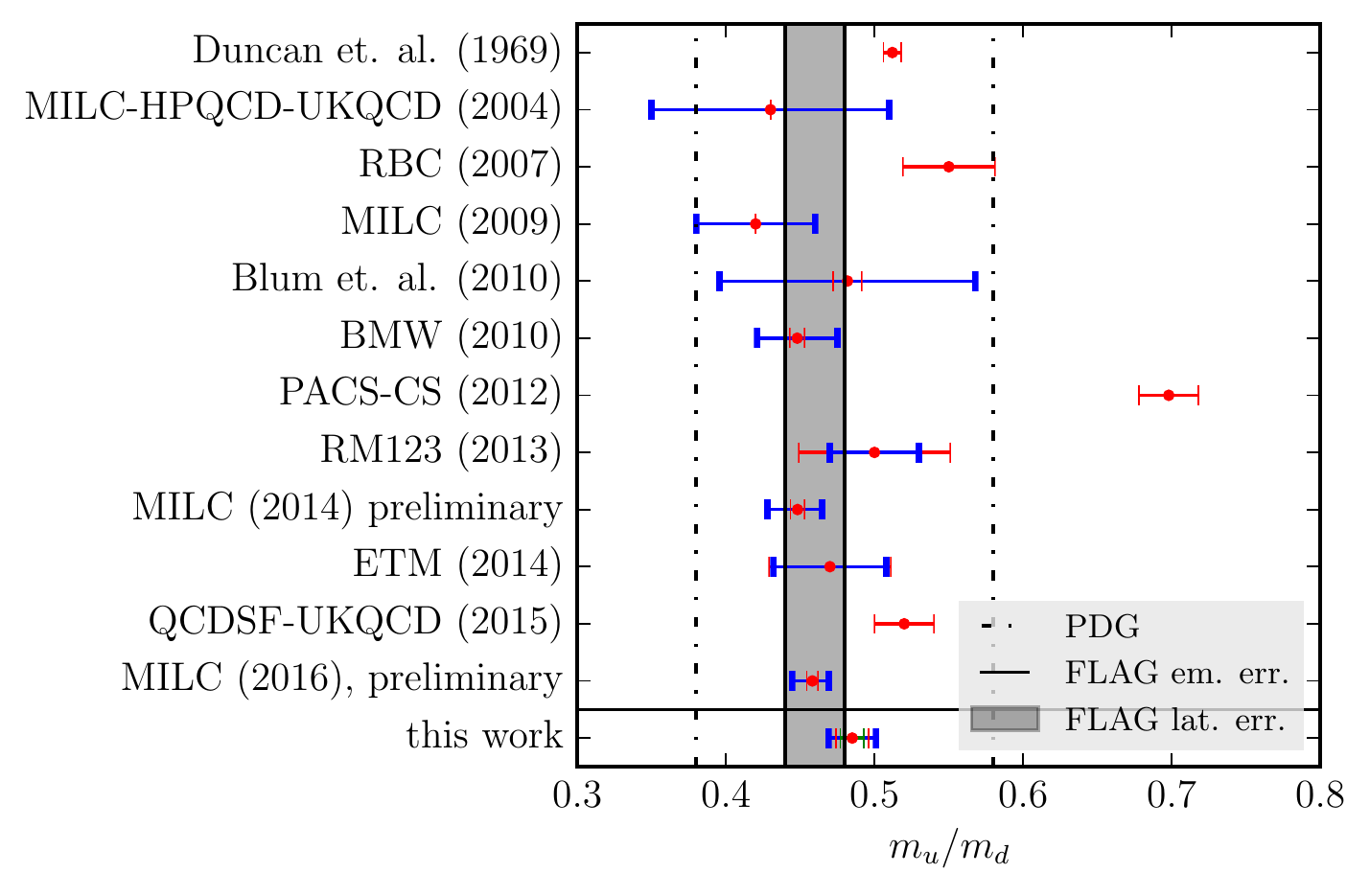}
\includegraphics[width=0.45\textwidth]{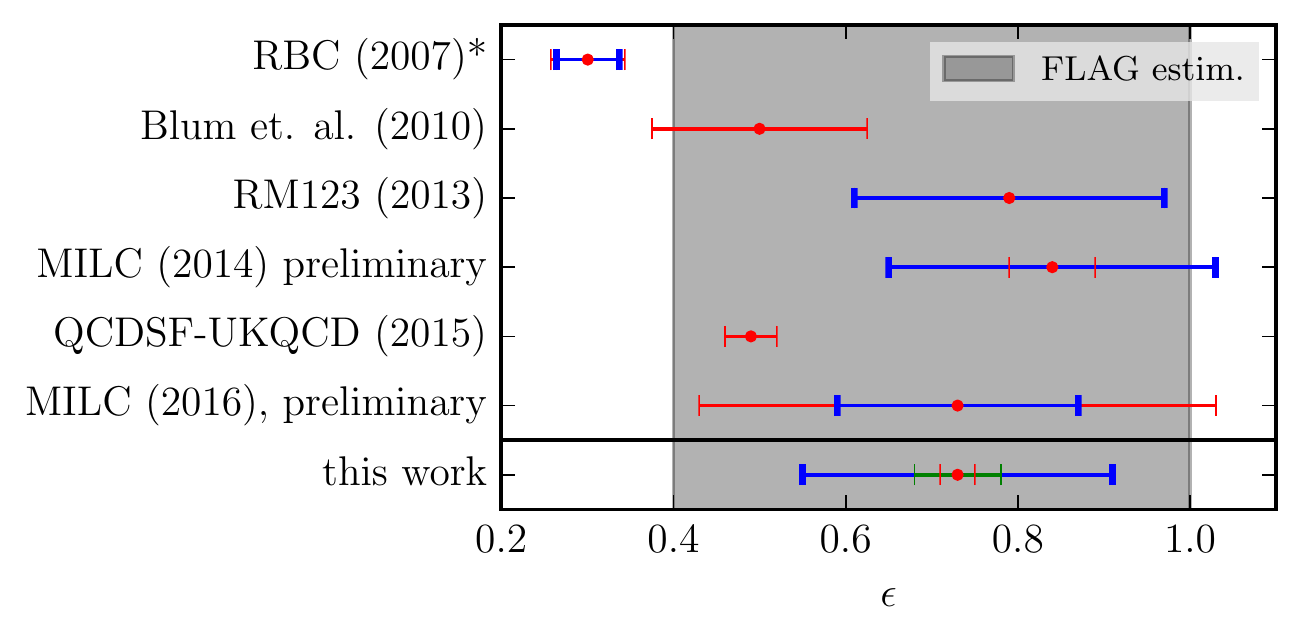}\\
\tiny{*: Systematic error: Difference of two results extracted from paper.}
\caption{Comparison between our values for the light quark mass ratio and the violation to Dashen's theorem to other determinations. Red error bars are statistical, blue error bars are systematic and blue error bars are our error without the QED quenching uncertainty. \cite{Durr:2010aw,Durr:2010vn,Aoki:2016frl,Bazavov:2009fk,Duncan:1996xy,Blum:2007cy,Carrasco:2014cwa,Aoki:2012st,Aubin:2004ck,deDivitiis:2011eh,Blum:2010ym,deDivitiis:2013xla,Horsley:2015vla,Basak:2016jnn,Olive:2016xmw}\label{fig_comp}}
\end{figure}

Using the fitted value of $\Delta_{\mbox{QED}} M_K^2 = C_K \alpha |_{\mbox{phys. pt.}}$ and the experimental Pion mass splitting one can compute using eqn. (\ref{eqn_eps}) the correction to Dashen's theorem.  Using $\Delta_{\mbox{QED}} M_\pi^2/\Delta M_\pi^2=0.04(2)$ from \cite{Aoki:2016frl} the small error introduced by replacing $\Delta_{\mbox{QED}} M_\pi^2$ by $\Delta M_\pi^2$ can be corrected resulting in $\epsilon_c$. The results are:
\begin{equation}
 \epsilon = 0.73(2)(5)(17) \ , \ \ \ \epsilon_c = 0.77(2)(5)(17)(2)
\end{equation}
where the first error is statistical, the second error is due to systematics uncertainties in our analysis and the third error is an estimate of the uncertainty introduced by QED quenching. The last error is due to the above mentioned correction.
Furthermore eqn. (\ref{eqn_del_m}) was used to infer
\begin{equation}
 \delta m = m_u - m_d = -2.41(6)(4)(9) \, \mbox{MeV}.
\end{equation}
When this value is combined with the previous result $m_{ud}=3.469(47)(48) \, \mbox{MeV}$ from \cite{Durr:2010vn} it can be derived that
\begin{equation}
 m_u = m_{ud} + \delta m/2 = 2.27(6)(5)(4) \, \mbox{MeV} \ , \ \ \
 m_d = m_{ud} - \delta m/2 = 4.67(6)(5)(4) \, \mbox{MeV} 
\end{equation}
in the $\overline{\mbox{MS}}$-scheme at $2 \, \mbox{GeV}$. Therefore the ratio of light quark masses is
\begin{equation}
 m_u/m_d = 0.485(11)(8)(14).
\end{equation}
While this ratio in principle is scale and scheme dependent this can be neglected at the leading order in the isospin splitting. It is also interesting to derive results for the flavor breaking ratios $R$ and $Q$:
\begin{equation}
 R = \frac{m_s-m_{ud}}{m_d - m_u} = 38.2(1.1)(0.8)(1.4) \ , \ \ \
 Q = \sqrt{\frac{m_s^2-m_{ud}^2}{m_d^2 -m_u^2}} = 23.4(0.4)(0.3)(0.4).
\end{equation}
In general the results are in good agreement with the FLAG estimates \cite{Aoki:2016frl}. A comparison of two results with the PDG value \cite{Olive:2016xmw}, the FLAG value \cite{Aoki:2016frl} and the references therein can be found in figure \ref{fig_comp}.

\section{Acknowledgement}
Computations were performed using the PRACE Research
Infrastructure resource JUGENE based in Germany at FZ
Jülich and HPC resources provided by the ``Grand
équipement national de calcul intensif''
(GENCI) at the ``Institut du développement et des
ressources en informatique  scientifique'' (IDRIS)
(GENCI-IDRIS  Grant No. 52275), as well as further
resources at FZ Jülich and clusters at Wuppertal
and Centre de Physique Théorique.
This work was supported in part by the OCEVU Laboratoire
d’excellence (ANR-11-LABX-0060) and the A*MIDEX
Project (ANR-11-IDEX-0001-02) which are funded by the
``Investissements d’Avenir'' French government program
and managed by the ``Agence nationale de la recherche''
(ANR), by CNRS Grants No. GDR No. 2921 and PICS
No. 4707, by EU Grants FP7/2007-2013/ERC No. 208740,
MRTN-CT-2006-035482 (FLAVIAnet), by DFG Grants
No. FO 502/2, SFB TRR-55 and UK STFC Grants
No. ST/L000296/1 and ST/L000458/1. L. V. was partially
supported by a GSI grant.

\end{document}